\documentclass[preprint,12pt]{elsarticle}

\usepackage{amssymb}

\usepackage{hyperref}

\journal{Journal of Physics: Condensed Matter}

\begin{document}

\begin{frontmatter}

\title{Magnetization and Thermal Entanglement of the Spin-1 Ising-Heisenberg Polymer Chain}

\author{A. Sadrolashrafi$^1$, V. Abgaryan$^2$, A. Rezvan$^3$, and N. S. Ananikian$^1$}

\address{$^1$ Alikhanyan National Science Laboratory, Alikhanian Br. 2, 0036 Yerevan, Armenia\\
$^2$ Laboratory of Information Technologies, Joint Institute for Nuclear Research, Joliot-Curie 6, 141980 Dubna, Moscow region, Russia 141980 Dubna, RU\\
$^3$ Erich Schmid Institute of Materials Science, Jahnstrasse 12, 8700 Leoben, Austria}

\fntext[label1]{afsaneh@mail.yerphi.am}

\fntext[label2]{vahagnab@gmail.com}

\fntext[label3]{amir.rezvan@oeaw.ac.at}

\fntext[label4]{ananik@yerphi.am}

\date{\today}

\begin{abstract}
We establish a solvable Heisenberg-Ising model on a spin-1 Ni-containing polymer chain, $[Ni (NN'-dmen) (\mu-N_3)_2]$, with $NN'-dmen$ being $NN'-dimethylethylenediamine$, that fully covers the interaction characteristics of the material and by which, we can characterize all the peculiar magnetic features of the polymer, which has been partly studied in experiment. By purely analytical calculations, we can see that the magnetization exhibits three plateaus at zero, mid, and 3/4 of the saturation value at low temperatures below 2 K. The corresponding featuring peaks of magnetic susceptibility are clearly shown. The model also displays plateaus in thermal entanglement that captures the one-to-one correspondence between thermal entanglement plateaus and those of the magnetization. The calculations are done by the transfer matrix technique.
\end{abstract}


\end{frontmatter}

\section{Introduction}
The $\mu$-azido bridging ligand with divalent metal ions, mainly $Cu^{II}$, $Ni^{II}$, $Co^{II}$, $Cd^{II}$, $Fe^{II}$, and $Mn^{II}$ \cite{Ribas99, Mautner 2015} is a very good candidate and one of the most adaptable and flexible ones for creating new materials with different magnetic properties. Besides, it is very functional in the studying of magnetization structure and magnetic correlations both in discrete and polymeric complexes. The nitrogen $\mu$-azido may give end-to-end (EE) or end-on (EO) coordination modes, where normally, the first ones cause antiferromagnetic couplings and the latter ones result in ferromagnetic couplings\cite{Ribas96}. 

\begin{figure}[htb]
\centering
\begin{tabular}{@{}cc@{}}
\small{(a)}\\ \cr
    \includegraphics[scale=0.35]{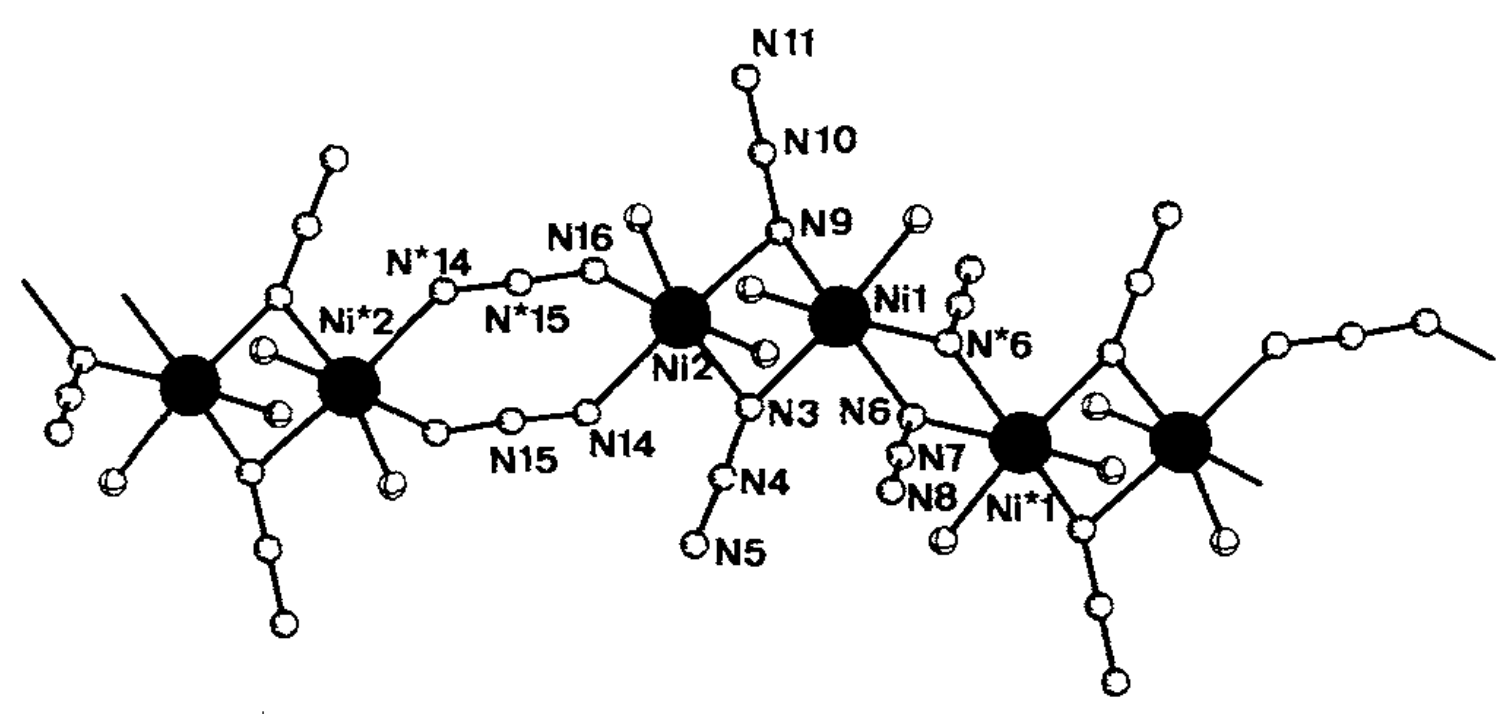} \\
\cr\cr \small{(b)}\\ \cr
    \includegraphics[scale=0.35]{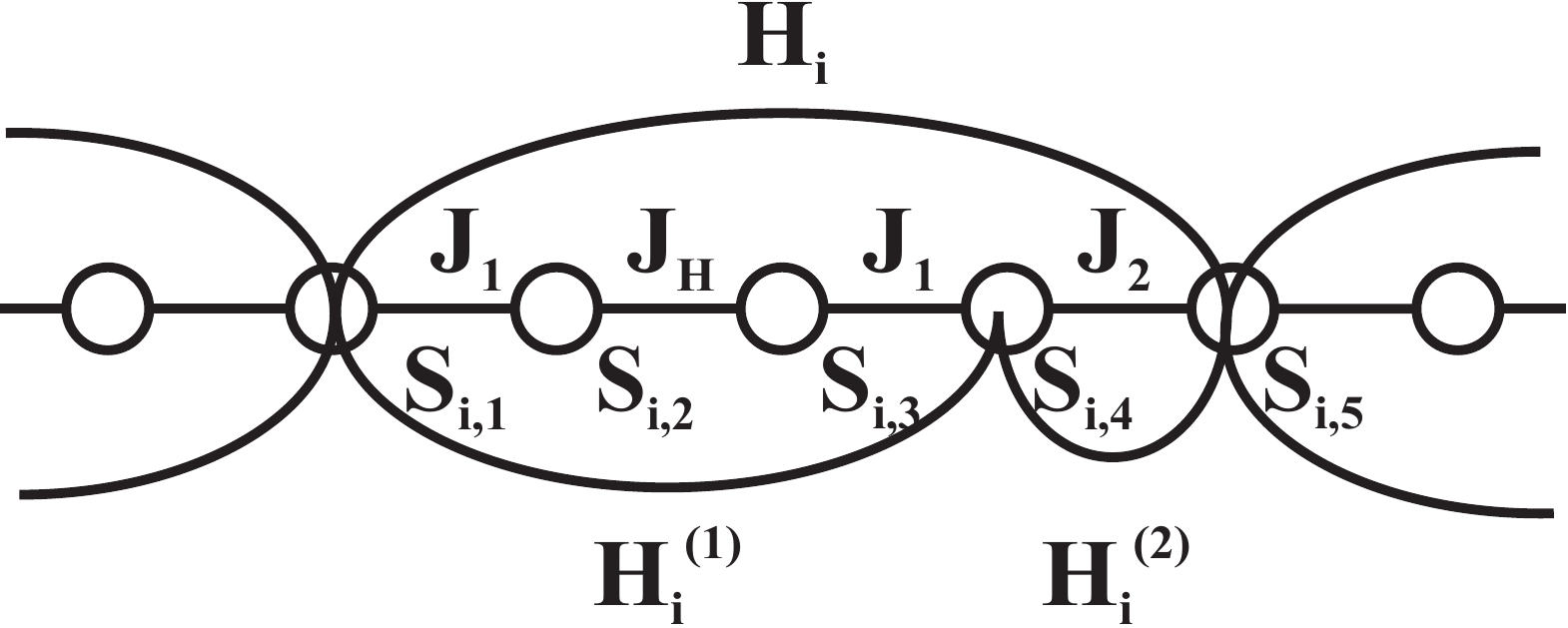}
\end{tabular}
\caption{(a) Spin-1 $Ni$-containing polymer $[Ni (NN'-dmen) (\mu-N_3)_{2}]$; See \cite{Ribas96} for more details on the molecular structure. (b) The coupling constants and the blocks composition in spin-1 $Ni$-containing polymer.}
\label{polymer}
\end{figure}

Joan Ribas et. al. in their experimental work \cite{Ribas96} reported that they synthesized and fully characterized a new $Ni^{II}$-compound, with chemical formula, $[Ni (NN'-dmen) (\mu-N_3)_2]$, where $NN'-dmen$ is $NN'-dimethylethylenediamine$ that exhibits the two kinds of coordination mode at the same time (figure \ref{polymer}(a)): the molecular structure consists of $Ni^{II}$ ions centers alternatively linked by three double EO entities and one double EE entity. Each $Ni^{II}$ ion completes its distorted octahedral coordination by binding to one bidentate $NN'-dmen$ ligand. 

Here in this paper, We report on the magnetic properties of the same polymer through introducing an exactly solvable model that truly describes the interaction characteristics of the spin-1 $Ni$-containing polymer \cite{Ribas96} at low temperatures. We also look into negativity as a measure to observe quantum correlations and entanglement of the polymer. 

\section{Solvable spin-1 model Hamiltonian}\label{Model}
The spin chain Hamiltonian $H = \sum_{i=1}^{N} H_{i} $ with
\begin{eqnarray}\label{Tot Ham}
\nonumber H_{i} &=& H_{i}^{(1)} + H_{i}^{(2)}\\ 
\nonumber H_{i}^{(1)} &=& - J_H \vec{S}_{i,2} . \vec{S}_{i,3} - J_{1}( S_{i,1}^z S_{i,2}^z + S_{i,3}^z S_{i,4}^z )\\
\nonumber &&- g \mu_{B} h ( \frac{1}{2} S_{i,1}^z + S_{i,2}^z + S_{i,3}^z + \frac{1}{2} S_{i,4}^z )\\
\nonumber &&+ D ( \frac{1}{2} (S_{i,1}^z)^2 +(S_{i,2}^z)^2+ (S_{i,3}^z)^2 + \frac{1}{2}(S_{i,4}^z)^2 )\\
\nonumber H_{i}^{(2)} &=& - J_{2} S_{i,4}^z S_{i,5}^z - \frac{1}{2} ( g \mu_{B} h ( S_{i,4}^z + S_{i,5}^z )\\
&&- D ( (S_{i,4}^z)^2 + (S_{i,5}^z)^2) )
\end{eqnarray}
captures the interaction structure in the polymer $[Ni (NN'-dmen) (\mu-N_3)_2]$, which is experimentally studied in \cite{Ribas96} and is a highly unusual one-dimensional polymer containing at the same time both kinds of coordination mode (EE and EO) and showing spectacular magnetic properties (figure \ref{polymer} (a)). In equation (\ref{Tot Ham}), $\vec{S}_{i,k}$ is the vector operator of spin-1 at site $k$ of the $i$-th block and $S_{i,k}^{z}$ is its corresponding $z$-component. Cyclic boundary conditions are applied in the thermodynamic limit. From the experimental data \cite{Ribas99, Mautner 2015, Ribas96}, we consider the approximate values of $J_1 = 20 cm^{-1}$, $J_2 = 37 cm^{-1}$, $J_{H} = -120 cm^{-1}$, and $D = -6 cm^{-1}$, where $J_1$ and $J_2$ are bilinear Ising ferromagnetic coupling constants, $J_{H}$ is Heisenberg antiferromagnetic exchange coupling, and $D$ is single-ion anisotropy. Also, h is external magnetic field, while $g = 2.39$ and $\mu_{B}$ are the Land\'{e} factor and the Bohr magneton respectively. Our exact calculations based on the Hamiltonian $H$, which is the spin-1 Heisenberg-Ising model with small bilinear Ising ferromagnetic coupling constants and rather large Heisenberg antiferromagnetic exchange coupling completely matches experimental magnetic results reported in the literature \cite{Ribas96}.

We see that the model of the polymer is actually composed of $N$ separable blocks each of which consists of 5 spins and where the first and 5th spins of each block is shared by the neighboring blocks (figure \ref{polymer}(b)). In our calculations based on the transfer matrix method, we utilize the following relations $Z = tr \prod_{i=1}^{N} e^{ - \beta H_{i} }$, $F = - \frac{1}{ \beta } \log{Z} \sim - \frac{N}{ \beta } \log{\lambda_{Max}}$, $m = - \frac{1}{ N } \frac{ \partial F }{ \partial h }$, and $\chi = \frac{ \partial m }{ \partial h }$, where $Z$ is the partition function, $F$ is the free energy, $m$ is the magnetization per site, and $\chi$ is the magnetic susceptibility. Also, $\beta$ is the inverse temperature $\frac{1}{k_B T}$, where $k_B \approx 0.695 \ cm^{-1} K^{-1}$ is the Boltzman constant and $\lambda_{Max}$ is the largest eigenvalue of the transfer matrix $W$ generated by the block Hamiltonian $H_{i}$, while $Z = tr W^N$ and the approximation $Z = \lambda_{Max}^N$ holds for sufficiently large N's. 

We can make the same calculations on the basis of also double-block computation approach with generating sub-Hamiltonians $H^{(1)}_{i}$ and $H^{(2)}_{i}$ in which, we have a block of four spins followed by a block of two spins, where the first and last spins of each block are always shared with the neighboring blocks (See figure \ref{polymer}(b).) and $W=W^{(1)}W^{(2)}$. The two approaches would obviously give the same results. In our calculations, on the basis of transfer matrix for the blocks, we would face a not very common case in which the transfer matrices of the two sub-blocks are symmetric while the product of them which gives the transfer matrix of the whole block of 5 particles is non-symmetric positive (See \cite{TM}, for example.). As it is well known, $W$ would be similar to a matrix in Jordan canonical form. Hence $tr W = tr J$, where $J$ is the Jordan form of $W$. Given that the transfer matrix is a matrix of positive elements, according to Perron-Frobenius theorem, we know that the eigenvalue with the largest absolute value would be real.

\section{Magnetic properties of the polymer}
\begin{figure}[htb]
\centering
\begin{tabular}{@{}cc@{}}
    \includegraphics[scale=0.3]{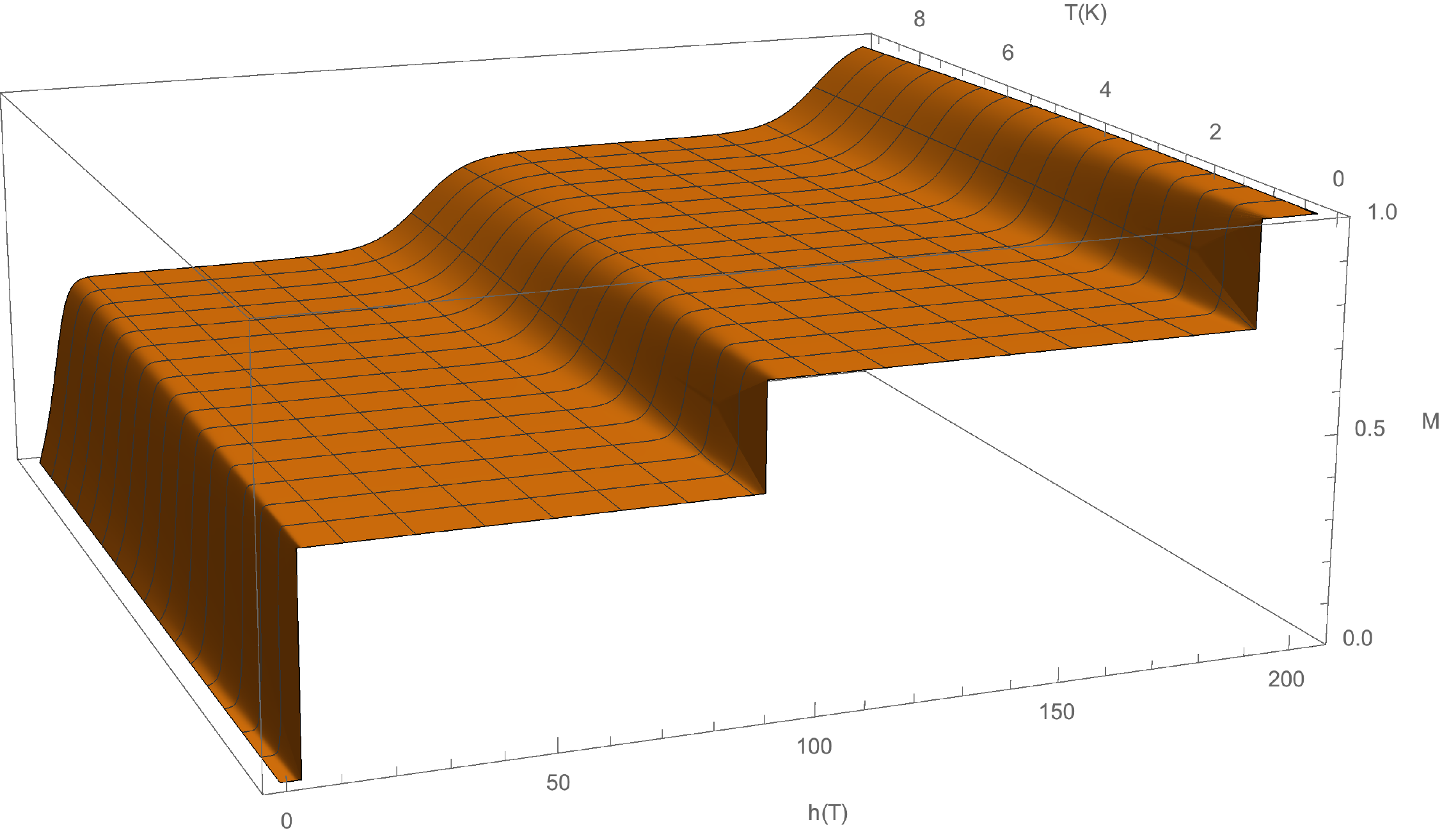} \\ \cr\cr
    \includegraphics[scale=0.27]{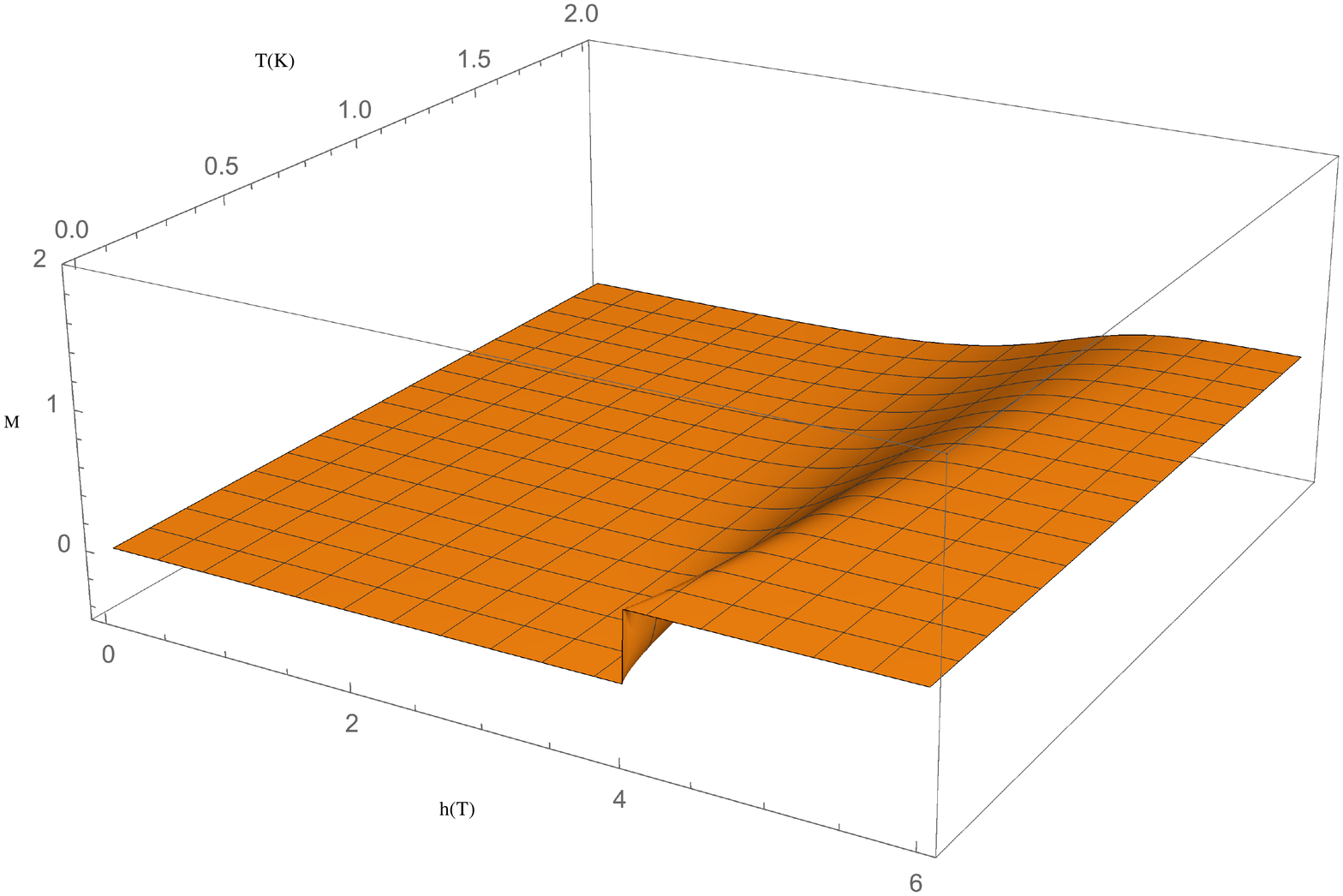}
  \end{tabular}
 \caption{Above: magnetization behavior of the spin-1 $[Ni (NN'-dmen) (\mu-N_3)_2]$ polymer as a function of the absolute temperature $T$ in $Kelvin$ and the magnetic field $h$ in $Tesla$ Below: the zero- and 1/2-plateau of the magnetization is showing. Magnetization has no units.}
\label{m}
\end{figure}

Our calculations show that for non-negative values of external magnetic field and at low temperatures, we observe three magnetization plateaus at $0$, $\frac{1}{2}$, and $\frac{3}{4}$ of the saturation value, which are clearly shown in Fig (\ref{m}) versus the applied magnetic field $h$ and the temperature $T$. Fig (\ref{m}) also shows the first plateau along with part of the second one for values of $0$ to $6$ $Tesla$ of the applied magnetic field and in temperatures below $2 K$.

Magnetization plateaus occur both in antiferromagnetic and ferromagnetic materials and they play an essential role in understanding a large family of nontrivial quantum phenomena. The phenomenon of magnetization plateau is considered as a macroscopic manifestation of the essentially quantum effect in which the magnetization $m$ is quantized at fractional values of the saturation magnetization $m_s$. The quantum plateau state was actually first discovered over two decades ago \cite{Hida}. One of the first experimental observations was reported in a diamond chain \cite{diamond chain exp 1}. Magnetization plateau has been studied during the past decade both experimentally and theoretically in spin-1 models \cite{diamond chain exp 2, diamond chain theor 2, Ananikian diamond chain3, Ananikian diamond chain4, india}. 

\begin{figure}[htb]
\centering
\begin{tabular}{@{}cc@{}}
    \includegraphics[width=.27\textwidth]{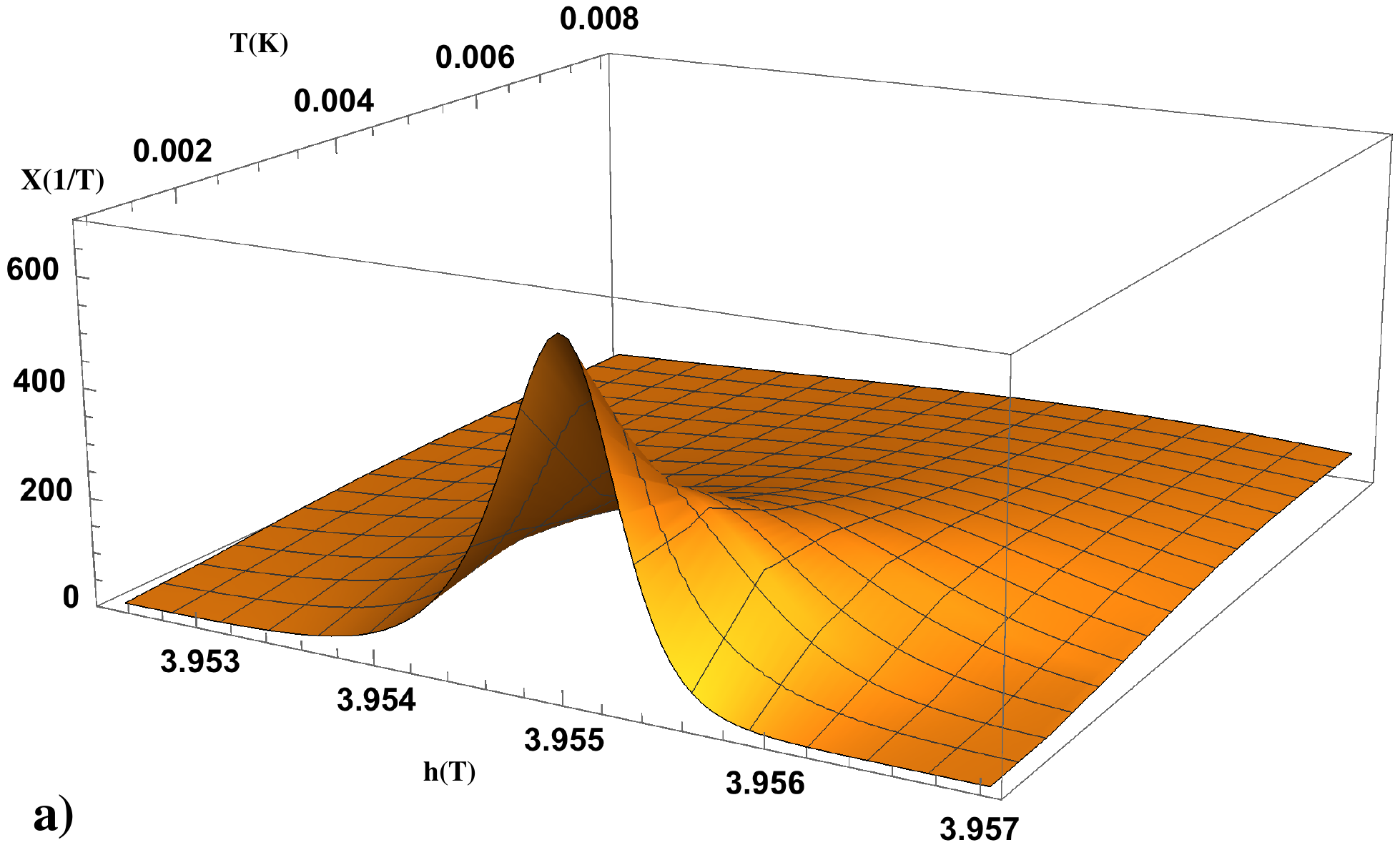} &
    \includegraphics[width=.27\textwidth]{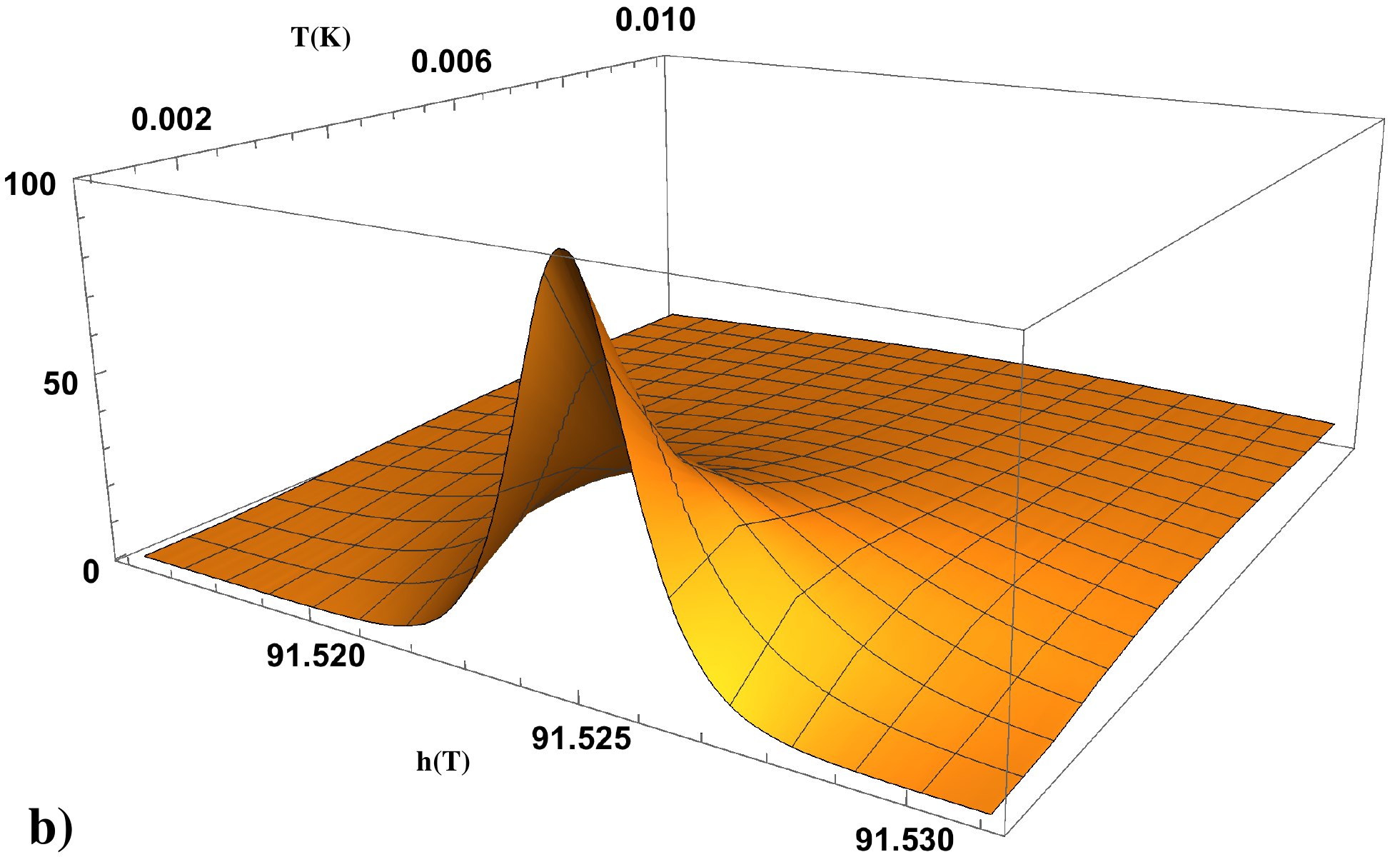}\\  
     \multicolumn{2}{c}{\includegraphics[width=.27\textwidth]{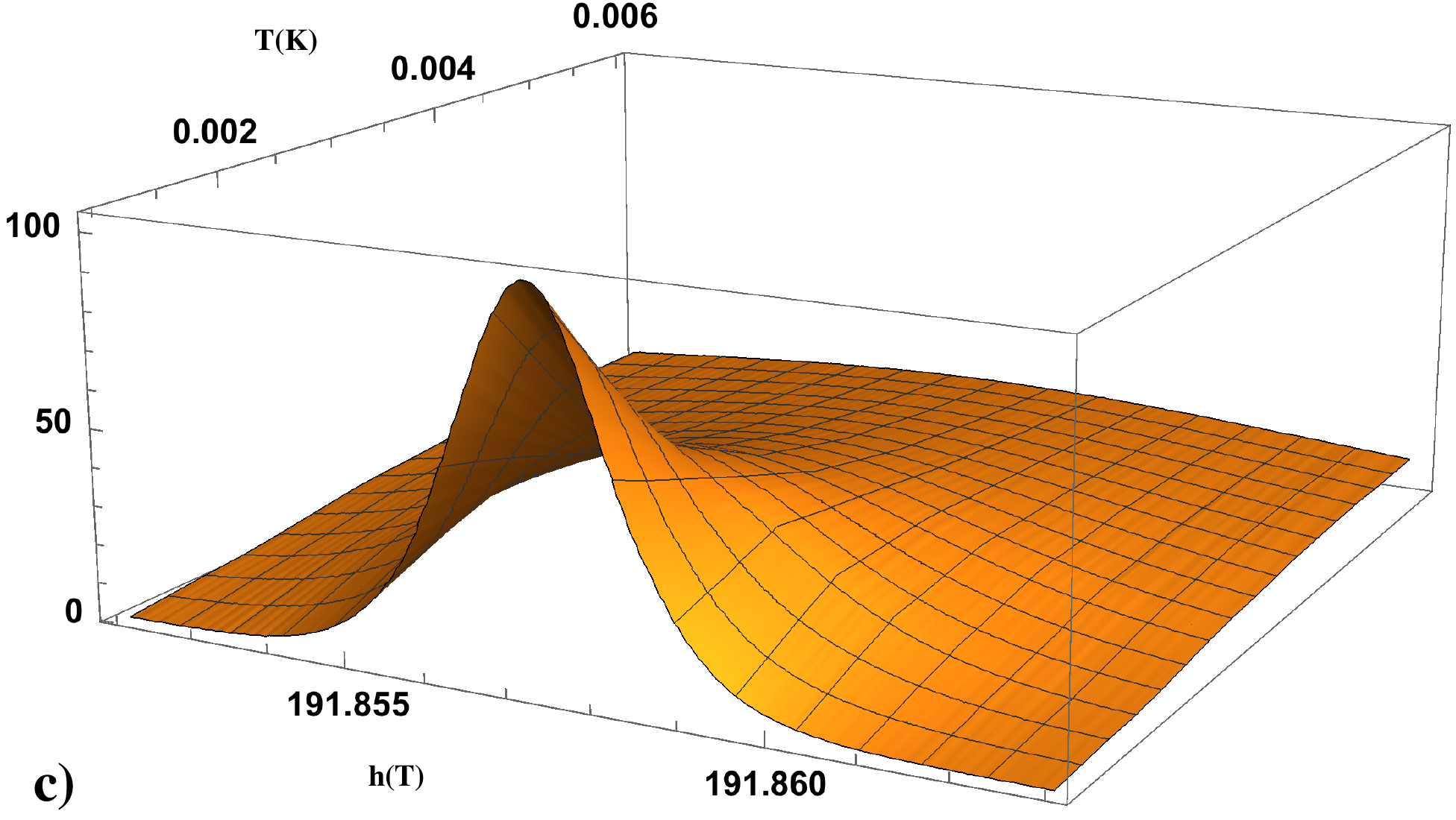}}
  \end{tabular}
\caption{The three peaks of the magnetic susceptibility of the spin-1 $[Ni (NN'-dmen) (\mu-N_3)_2]$ polymer as a function of the absolute temperature $T(K)$ and the magnetic field $h(T)$ at low temperatures. Magnetic susceptibility $\chi$ is measured in $cm^{-1}$.}
\label{suscep}
\end{figure}

In accordance with the appearance of three plateaus in the magnetization graph, we observe three corresponding peaks in the magnetic susceptibility which are shown in the Fig. (\ref{suscep}) as a function of the applied magnetic field $h$ and the temperature $T$.

The magnetic susceptibility of a material is equal to the ratio of the magnetization $m$ within the material to the applied magnetic field strength $h$. This ratio, strictly speaking, is the volume susceptibility, because magnetization essentially involves a certain measure of magnetism (dipole moment) per unit volume. In figure (4) of the reference \cite{Ribas96} the magnetic susceptibility is calculated in $cm^3/mol$. For single $Ni$ - containing polymer with cyclic boundary condition in the thermodynamic limit, the magnetization $m$ is dimensionless and the magnetic susceptibility $\chi$ has the dimension of inverse energy $[cm]$ \cite{Baxter}. The characteristic peaks of the magnetic susceptibility can be seen in the low temperatures.

\section{Thermal negativity of $Ni$-containing polymer}
Entanglement is a type of correlation that is quantum mechanical in nature. Studying entanglement in condensed matter systems is of great interest due to the fact that some behaviors of such systems can most probably only be explained with the aid of entanglement. The magnetic susceptibility at low temperatures, quantum phase transitions, chemical reactions are examples where the entanglement and correlation functions are the key ingredients for a complete understanding of the system \cite{Entanglement1, Entanglement2, Magnetic susceptibility - entanglement witness}. Furthermore, in order to produce a quantum processor, the entanglement in condensed matter systems becomes an essential concept. On the other hand, in systems like some molecular magnetic materials, the magnetic susceptibility can be directly related to an entanglement witness (EW)\cite{Entanglement3}. Thermal negativity and magnetization plateaus of spin-1 particles in a diamond chain Ising - Heisenberg model has been recently studied in \cite{diamond chain theor 2, diamond chain - neg}.

\begin{figure}[htbp]
\begin{center}
\includegraphics[scale=0.35]{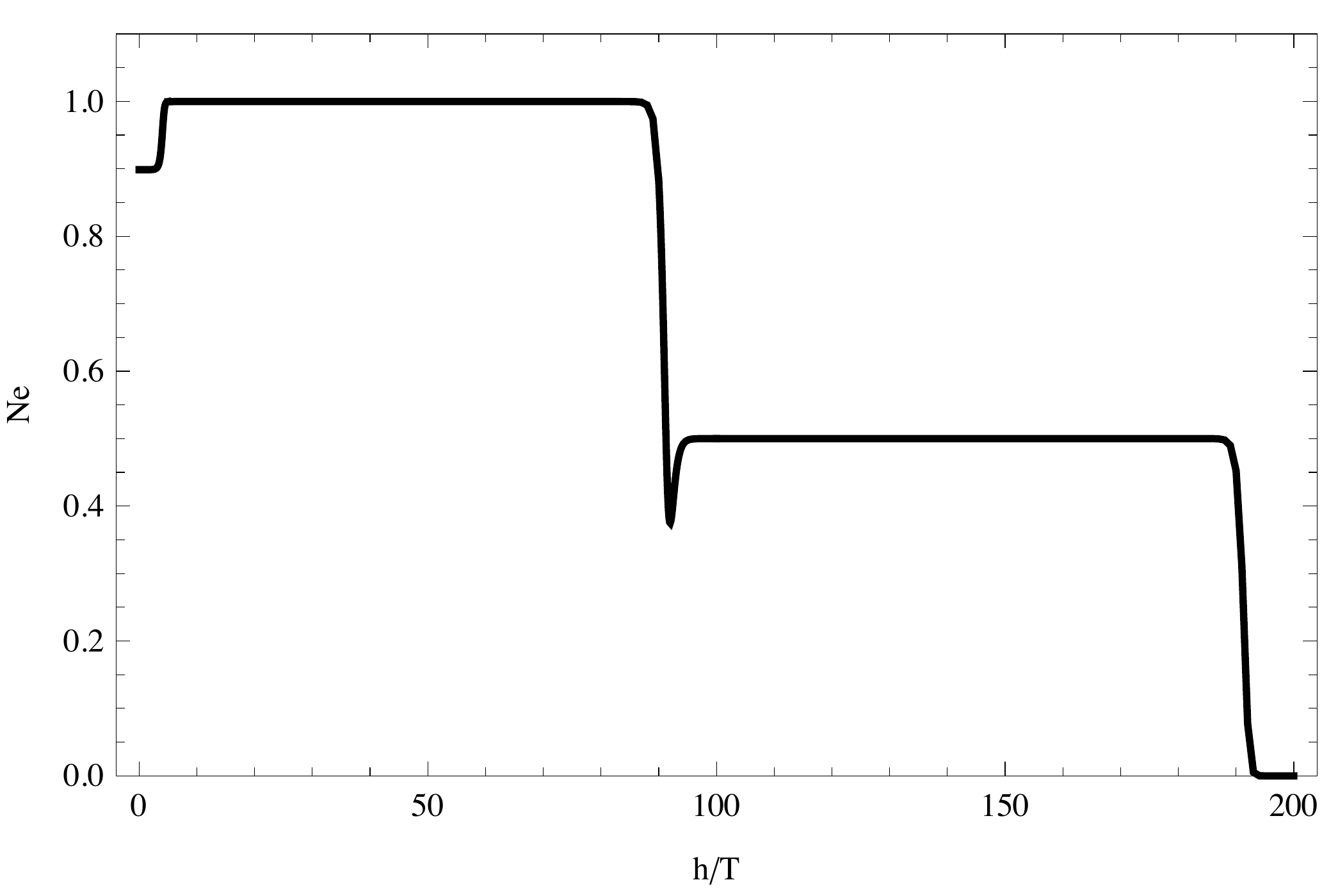}
\caption{the plateaus of the negativity in correspondence with the magnetization plateaus}
\label{n}
\end{center}
\end{figure}

We calculated negativity as a calculable measure of thermal entanglement \cite{neg} in Ni-containing $[Ni (NN'-dmen) (\mu-N_3)_2]$ polymer of spin-1 particles with Heisenberg-Ising interactions. For that purpose, we used the reduced density matrix $\rho$ of the pair of particles with Heisenberg interaction between them in the block. To find out the reduced density matrix $\rho$ in the $r$th block, we need to trace out all the degrees of freedom except for the 2nd and 3rd spin in block $r$. To accomplish such a calculation, we followed a method similar to the one that is described in \cite{diamond chain theor 2} through which, the reduced density matrix $\rho$ is derived in terms of the transfer matrix. Then the amount of entanglement between the two particles would be given by $Ne = \frac{ || \rho^{T_A} ||_1 - 1 }{2}$, where $\rho^{T_A}$ is the partial transpose of $\rho$ with respect to any one of its subsystems meaning any one of the spins coupled with antiferroagnetic Heisenberg exchange, whose elements are $\langle \alpha \gamma \vert \rho^{T_A} | \beta \delta \rangle = \langle \beta \gamma | \rho \vert \alpha \delta \rangle$ and $|| X ||_1$ is the trace norm of $X$, which is $Tr \sqrt{X^{\dagger} X}$ by definition. The result is showing in figure \ref{n}. We can see the one to one correspondence between the negativity plateaus and those of the magnetization in figure \ref{m}. 

\section{Conclusion}\label{conclusion}
In this letter we introduce and explore a compatible theory model of a Nickel-containing polymer that can fully cover and explain the experimental data which was gained and reported in \cite{Ribas96}. Along with studying magnetic properties of the model, we investigate quantum entanglement by calculating negativity for the Heisenberg-interacting pair in the model and we observe that it shows very tuned correspondence with the magnetization at low temperatures. We conducted our calculations on the basis of the transfer matrix method for separable blocks, where each two consecutive blocks would have their first and last spins in common. This study would enhance our understanding of nitrogen $\mu$-azido ligand as one of the most adaptable key entities that can be exploited to create new materials both discrete and polymeric and plays a key role in determining the relative magnetic properties of the materials. 
\section*{Acknowledgments}
N. A. acknowledges financial support by the CS MES RA  in the frame of the Research Project no. SCS 15T-1C114 and ICTP NT-04 grants. A. S. acknowledges the financial support from the fellowship granted by ICTP Office of External Activities (OEA), at the ICTP affiliated center at Yerevan, Armenia within NET68 and OEA-AC-100 programs.
\bibliographystyle{elsarticle-harv} 

\end{document}